%% file: main.tex
\documentclass[11pt]{article}
\usepackage{graphicx} 
\usepackage{booktabs}
\usepackage{tabularx}
\usepackage{caption}
\usepackage{hyperref}
\usepackage{doi}
\graphicspath{ {./images/} }
\usepackage{url}
\usepackage{authblk}
\providecommand{\keywords}[1]{\noindent \textbf{\textit{Keywords}} #1}

\title{\centering Informative Keyboard and its Application \\
to Raise Awareness of Smartphone Use}
\author{Jaroslaw Domaszewicz}
\author{Damian Sienicki}
\author{Michal Obirek}
\affil{\centerline{Institute of Telecommunications and Cybersecurity} \centerline{Warsaw University of Technology} \centerline{Nowowiejska 15/19, 00-665 Warsaw, Poland}}
\date{}

\begin{document}
\maketitle
\begin{abstract}
    \noindent Excessive smartphone use is now widely considered a personal and societal problem. It is recognized by application and smartphone makers, who provide tools to track the amount of use, set limits, or block certain services at predefined times. These tools, while powerful, may require significant cognitive effort to operate: configuration parameters need to be set, and captured statistics need to be analyzed. To offer a complementary solution, we propose a radically different approach. We employ the keyboard of a smartphone as an output device. With each press of a key, the user is given a high-level, qualitative, color-encoded estimate of the amount of recent smartphone use. The technique, dubbed the informative keyboard, is a case of implicit interaction: the user's intention is to enter text but, while typing, they receive the feedback. In the paper, we elaborate the concept, identify design decisions, describe our implementation, present the outcome of a questionnaire-based evaluation, and point to some other applications of the informative keyboard.
\end{abstract}



\input{./introduction}
\input{./overview}
\input{./related_HCI}
\input{./heating_up}
\input{./survey}
\input{./related_work}
\input{./summary}


 \section*{Acknowledgements}

 The authors thank Warsaw University of Technology (WUT) students Michał Banasik and Julian Uziembło for their contribution to the implementation of the informative keyboard. We thank the students of the course ``IoT and Network Programming,'' taught at WUT, for participating in the evaluation session and providing the insightful feedback. We also appreciate instructive discussions with Dariusz Parzych of the Strategic Analysis Department at WUT.

\bibliographystyle{IEEEtranDOI}
\bibliography{keyboard}
\end{document}

%% file: introduction.tex
\section{Introduction}
The massive usage of smartphones by people of any age, gender, profession, or country can be easily observed in most daily situations. In \cite{clayton2022uk}, it is reported that as many as 44\% of the people living in the UK look at a smartphone screen for at least four hours a day, and 20\% of the people look for at least six hours (as of 2022). Smartphone overuse and its possible impact on the individual and society have attracted a lot of attention. For example, in \cite{carr2017how}, the author concludes that smartphones impair our ability to think, interact with people directly, and remember meaningful things. The negative impact is experimentally confirmed in \cite{ward2017brain}, which shows that simply placing a smartphone on the desk while working reduces the cognitive capacity of the owner, even if the device does not produce any notifications. In \cite{wacks2021excessive}, the authors review evidence linking excessive smartphone use to psychological and medical problems in young people. We take it for granted that many people would like to better control the amount of time they spend with their smartphones, in order to improve their digital wellbeing.

The problem of excessive smartphone use has been recognized by smartphone makers, application developers, and researchers. There exist many digital self-control tools (DSCTs) \cite{roffarello2023achieving}, some of which offer extensive, detail-oriented functionality. Consider Digital Wellbeing \cite{googleWWWdigital}, the flagship Android application developed to help people take control. It features: (a) a bedtime mode, when the screen gets gray and notifications are silenced, (b) two profiles, for work and home, (c) turning off interruptions by putting the device face down, (d) a focus mode, when selected distracting apps are paused, (e) easy muting of all notifications and calls, (f) picking apps that can send notifications, (g) setting per-app timers, which, once expired, make an application pause, and (h) a dashboard that displays a per-app breakdown of the time of smartphone use.  

Configuring such detail-oriented solutions may require defining schedules, selecting distracting applications, or setting per-application time limits. Then, assorted usage statistics are regularly offered for inspection. Often, the user interface consists of several screens. The mental effort needed to operate an application of this kind can paradoxically add to smartphone-related overload rather than reduce it. In addition, some parameters (for example, the bedtime schedule), when not adjusted by context of use, may prove inadequate under specific circumstances. These complexities may lead users to abandon efforts to monitor and control how they use their smartphone.

While not disputing the utility of applications like Digital Wellbeing, we adopt a radically different approach to smartphone use monitoring. It is meant to complement the approach based on multiple settings and detailed feedback. We aim to produce minimal feedback that amounts to a high-level, one-item summary. The feedback should be provided implicitly, without any user action required. The feedback should be nonintrusive but at the same time hard to overlook. The feedback should be delivered in quasi-real time, roughly concurrently with the usage itself. We want to represent the feedback without text, to make it easy to absorb. The overall guiding principle is to gently but firmly push the user to become aware of how much they use their smartphone, with minimal extra cognitive load and without relying on continued intention to monitor the usage. We aim at awareness that comes from self-monitoring, without blocking anything or restricting the user otherwise. Some existing digital self-control tools were developed along similar lines; however, to the best of our knowledge, our approach is unique.

To achieve the above objectives, we propose two mechanisms: one that produces the minimal feedback and one that delivers it to the user. As to the former, a one-number running estimate of the amount of smartphone use over some recent time (say, several hours) is maintained. The estimate may be consolidated, combining all uses, or it may capture one specific way of using a smartphone, such as opening a selected application. The estimate is updated (sampled) at regular intervals, on the order of half an hour. Finally, the estimate is quantized to just a few levels, to make it easy to deliver and interpret. 

For the delivery mechanism, we propose what we call an "informative keyboard." While the user is typing, the informative keyboard serves as both an input device and an output device. The output is realized via colored pop-ups on keypress. Each level of the estimate is assigned its own color, and the pop-ups are colored according to the current level. The informative keyboard delivers information implicitly, without any additional user actions; the delivery is piggybacked on typing, which remains \textit{the} reason to use the keyboard. The sampling frequency of the usage estimate is such that its level (and the pop-up color) likely remains the same throughout a typing session. Thus, the user can keep the color on the periphery of attention. At the same time, the feedback can hardly remain unnoticed; each time the user presses a key, they receive a separate, discrete, pulse-like stimulus, which encodes a level of the amount of smartphone use. As typing occurs multiple times a day, estimate levels are delivered just as often. While some hours may pass without typing, in most cases the feedback is delivered frequently enough, so that the user can adjust their behavior on the very same day (this is what we mean by delivery in quasi-real time).  

This paper makes the following contributions. First, we introduce the informative keyboard, i.e., a keyboard that delivers information while the user is typing. It is conceived as an output device that can work with different sources of information. Second, we propose to use the informative keyboard to raise the user's awareness of the amount of their smartphone use, without asking for much attention or mental effort. We also present one way to derive minimal feedback. Third, we briefly describe our proof-of-concept implementation of the informative keyboard. Fourth, we present the results of an evaluation session, in which our concept and implementation of the informative keyboard were demonstrated to sixty seven university students. 

The paper is organized as follows. In Section \ref{section:overview}, we explain how the informative keyboard works. In Section \ref{section:HCI}, we position the informative keyboard against the background of related concepts in Human-Computer Interaction (HCI). In Section \ref{section:heating}, we narrow down our focus to feedback on smartphone use. There, we cover a possible approach to derive minimal feedback, and we present our proof-of-concept implementation of the informative keyboard. In Section \ref{section:survey}, we report on the results of the evaluation session. In Section \ref{section:related_DSCT}, we mention digital self-control tools that share some features with our solution. We summarize the paper in Section \ref{section:summary}, where we also comment on how to further explore the informative keyboard, enhance it, and apply it in different scenarios.

%% file: overview.tex
\section{Informative keyboard: overview} \label{section:overview}

Pop-ups on keypress is a popular feature of smartphone keyboards. If enabled, each time a key is pressed, an enlarged image of the key (a pop-up) is displayed. The user can glance at the pop-up to make sure that they have pressed the right key. The pop-up, rectangular or circular, contains the character of the key against a neutral (e.g., gray) background.

\begin{figure}[t]
\centering
\includegraphics[scale=0.7]{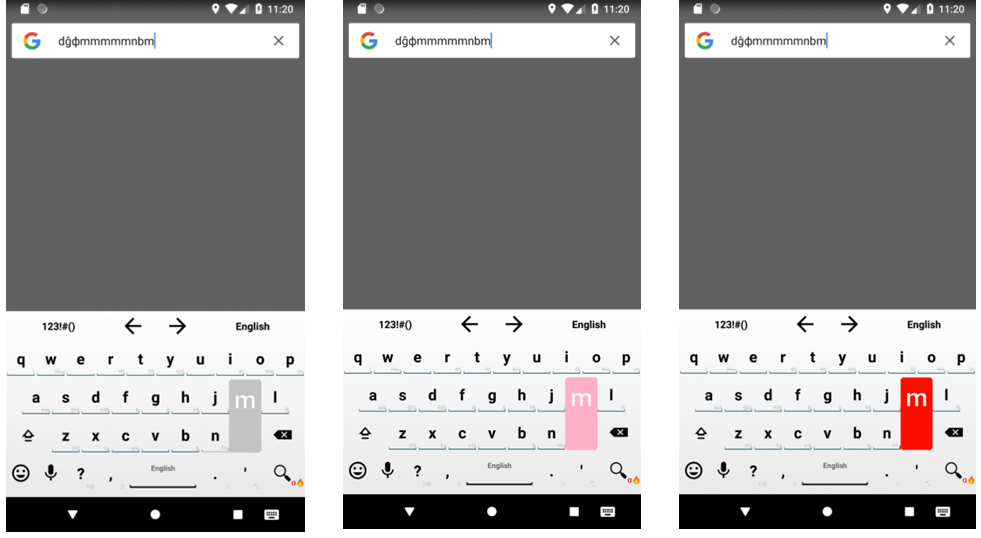}
\caption{The informative keyboard delivers information by changing the color of the background of pop-ups on keypresses. The user interprets the color in terms of a message. In the figure, three different messages are delivered at different times. Messages arrive at a low rate, so the color changes infrequently (there are many consecutive, same-color pop-ups). The messages are non-critical. }
\label{fig:informative}
\end{figure}

The informative keyboard is first and foremost used to type, just like a regular one. However, while being used to type, it also provides information. This is done using pop-ups on keypress: when a key is pressed, information is delivered via the color of the pop-up background. Each possible color has its own meaning and represents a unique piece of information. The basic principle of operation is presented in Figure \ref{fig:informative}. We refer to this mechanism as the output channel of the informative keyboard.

The output channel is driven by a source of information. The source is discrete and produces a sequence of ``messages'' from a finite ``alphabet.'' If the information is originally analog, it may be sampled and quantized, i.e., mapped to a finite set of quantization levels. Messages are produced at discrete time instants, which need not be uniformly spaced. The sequence of messages and their timing depend solely on the source; they are not affected by the keyboard. We conceptually decouple the source of information and the informative keyboard as an output device.

At any given time, the pop-ups for all keyboard keys use the same color, the one corresponding to the most recent message (referred to as the current message); thus, it does not matter which keys are pressed. The color changes when the source produces a new message. It is assumed that the user knows what messages the source can produce (i.e., the alphabet), and how they are encoded with color. As a result, they can interpret the background color of a pop-up as a specific message.

To make it easy to recognize colors, we focus on sources with tiny alphabets of no more than eight messages, and preferably around five. We assume that the messages are produced at a low rate (say, a couple per hour). Thus, when typing, the user usually receives the current message on multiple consecutive keypresses, i.e., successive pop-ups have the same background color. We also assume that messages are not critical and can be missed without serious consequences. The assumption is needed because it is possible for a message to never get delivered; this happens when another message is produced before the keyboard is used. 

In this paper, we apply the informative keyboard to deliver feedback on the amount of smartphone use. However, the keyboard is a ``general purpose'' output device, which can work with other information sources (subject to the above constraints). Examples include a current level of outside air pollution, a type of weather predicted for the next day, a running estimate of the time spent in the sedentary position (as estimated by the smartphone), the number of weeks remaining until an important deadline, a range of the number of steps made today, or even messages from a loved one, indicating their current mood (from ``very happy'' to ``very unhappy''). As can be seen, messages can inform, remind, or motivate.

For many sources, a message represents a number (or a range of numbers), but this need not be the case; a weather type source produces nonnumerical (qualitative) information. If messages are originally numerical, there are so few of them in the alphabet that numbers can be conveniently described with short phrases. For example, an air quality monitoring site \cite{airlyWWWair} provides an air quality index (a number), but also a phrase reflecting the current value (e.g., ``Poor air quality''). This is how we envision the informative keyboard to be used in most cases: having recognized a color as a message, the user thinks of it as a phrase, not a number. 

Key design choices concerning the informative keyboard have to do with its palette of pop-up background colors and the assignment of colors to messages. These choices affect comprehension (the ease of recognizing colors) and intrusiveness (a possible negative impact of delivering messages on typing, the keyboard’s primary functionality). Concerning comprehension, representing information exclusively through color can cause accessibility problems and is discouraged in some contexts \cite{W3CWWWaccessibility}. Sight impairments, such as color blindness, may hinder the reception of color-encoded messages. Such reservations can be addressed by drawing insights from the domain of data visualization \cite{yi2019how}, where color is a key means of conveying information. For example, with a small alphabet, the palette includes only a few colors; this makes it easier to tell them apart. Also, some palettes are better than others; for example, the colors should differ not only in hue, but also in lightness \cite{W3CWWWaccessibility}. Clearly, there should be different palettes to choose from or the possibility to customize the palette. The type of the palette should be in line with the nature of the information source: sequential palettes for numerical sources, and categorical (qualitative) palettes for nonnumerical ones \cite{yi2019how}. For some specific quantities, e.g., air pollution levels or weather temperature, well-established and widely recognized color mappings are available.

In the case of a desktop PC or laptop, an RGB keyboard (an LED backlit keyboard) offers the capability to set a color that backlights each individual key. As such, it can be transformed into an output device. We found one proposal that uses an RGB keyboard to deliver information \cite{kaitosaari2023using}. Our work differs in that (a) we focus on smartphones as opposed to desktops or laptops, and (b) we use pop-ups, which are a unique feature of smartphone keyboards. In addition, we cover our proposal in much more detail.


%% file: related_HCI.tex
\section{Informative keyboard: related HCI concepts} \label{section:HCI}

The output channel of the informative keyboard works in an unusual interaction mode. It is neither push-based, as it does not rely on notifications, nor pull-based, as it does not rely on the user’s action to retrieve messages. To better understand the interaction mode, in this section we characterize the informative keyboard by referring to highly relevant HCI (Human-Computer Interaction) concepts: implicit interaction, peripheral displays, and persuasive technologies.


\subsection{Informative keyboard as implicit interaction device}

The most common meaning of the term ``implicit interaction'' has to do with the lack of intention: implicit interaction occurs when a system responds in a way that ``goes beyond what the user has intended'' \cite{serim2019explicating}. Reacting to an input of the user, the system bundles an intended effect with an unintended one. The input and the intended effect form a primary (explicit) interaction, while the input and the unintended effect form an implicit interaction \cite{serim2019explicating}. The user of the informative keyboard intends to type, but when they type, both text is entered and messages are delivered (the intended and unintended effect, respectively). Interestingly, the user may start typing just to see the current message; this so-called co-option \cite{dix2002beyond} is not uncommon in interfaces with implicit interaction.

Another sense of ``implicit interaction'' is related to the user’s level of attention: implicit interaction occurs in the attentional background \cite{serim2019explicating}. We cover the issue of attention in the next subsection, where we view the informative keyboard as a peripheral display. Yet another meaning of the term has to do with the user’s awareness of an unintended system response \cite{serim2019explicating}. So-called incidental interaction \cite{dix2002beyond} occurs when the system delivers some unintended value, of which the user is not aware. In \cite{dix2002beyond}, a ``continuum'' with three levels of interaction is presented: intentional, expected, and incidental. In our case, typing is intentional, while message delivery is nonintentional but expected: the user expects that it will happen while typing and is aware of it when it happens.

Having established that the keyboard uses implicit interaction (in the first sense of the term), we now comment on different aspects of the implicitness. Message delivery is attached to a regularly performed activity (i.e., typing), so no new user routine is needed. In other words, the implicit interaction is linked to what the user ``would have to perform anyway'' for a well-established primary task \cite{serim2019explicating}. Typing itself is a human-computer interaction, this time an explicit and attention-demanding one. Combining implicit interaction with preexisting explicit interaction is a popular design approach \cite{serim2019explicating}.

Usually, implicit interaction is meant to support the user in achieving a primary task. If lights are automatically turned on when the user enters a room, the implicit response helps in the primary task of entering. In our case, the implicit interaction (i.e., message delivery) does not support the explicit interaction (i.e., typing). The latter is only a medium for the former. Such a bundle could be called piggybacked implicit interaction, to stress that the implicit part takes advantage of the explicit part but is not related to it. 

A common motivation for implicit interaction is to spare the user an action, e.g., the action of switching on lights when entering. The user of the informative keyboard is spared the effort of retrieving messages explicitly. However, the leading motivation to combine message delivery with typing is not to reduce effort but to make it very likely for the user to be exposed to messages. Typing is essential and performed regularly, so exposure to messages is likely or even certain. The typical user will type multiple times a day, to text or search. Evidence of the high intensity of keyboard use in smartphones can be found in \cite{druijff2021behavioural} and \cite{schmidt2000implicit}. There, time-stamped keypresses are used to detect the timing of sleep. In these studies, conducted with college students, the number of hours with some keyboard activity was not less than twelve and thirteen per day, on average \cite{druijff2021behavioural}, \cite{schmidt2000implicit}. With the informative keyboard, each typing activity leads to the delivery of a message. Deliveries are very likely, without relying on the user’s intention, or even when the user is reluctant to receive the underlying information. 

Admittedly, the delivery of messages via the informative keyboard is opportunistic and unreliable (best effort). It occurs only when the user decides to type: ``explicit interaction becomes a prerequisite for implicit interaction'' \cite{serim2019explicating}. A message can be delivered with substantial delay or not at all. However, this is less of a problem than it appears. First, the messages are not critical. Second, as argued above, the typical user will have the opportunity to receive multiple messages every day. 

A frequent feature of implicit interaction is that unintended effects are produced in a context-aware fashion, based on system-sensed conditions \cite{schmidt2000implicit}. As described, the informative keyboard is not context-aware: messages are delivered unconditionally, whenever the user is typing. In the final section, we list context awareness as a possible enhancement and suggest how context can be applied.


\subsection{Informative keyboard as peripheral display}

A peripheral display \cite{matthews2004toolkit} provides noncritical information without distracting the user; thus, it is possible to keep it at the periphery of attention. (If a peripheral display is a distinct object in the environment, it is also called an ambient display.) The informative keyboard is designed so that typing should be at the center of attention, while message delivery should be at the periphery. Indeed, new messages arrive at a low rate, so one message is delivered many times (with consecutive keypresses); due to such redundancy, there is no need to pay attention to the pop-up on every keypress. In fact, since messages are not critical, the user may pay hardly any attention to the keyboard output channel. In addition, the informative keyboard has some features commonly found in peripheral displays: abstract message representations and low information capacity. In the following, we elaborate on two attention-related design dimensions identified for peripheral displays: notification level and intrusiveness \cite{matthews2004toolkit}, \cite{pousman2006taxonomy}, \cite{shelton2021gauging}. 

Peripheral displays may provide information at different notification levels \cite{pousman2006taxonomy}. The higher the notification level, the more attention is asked of the user \cite{matthews2004toolkit}. The informative keyboard, when not used, does not display anything and does not attract attention. The ``ignore'' notification level from the taxonomy proposed in \cite{matthews2004toolkit} seems to fit best; in \cite{pousman2006taxonomy}, it is called the ``user poll'' notification level. However, the keyboard does not rely on polling; instead, the user's decision to type is an opportunity to deliver a message. Thus, we propose a new notification level, ``wait for (user-induced) opportunity,'' a variety of ``ignore.'' 

When the user is typing, the notification level changes. Usually, the user receives many consecutive pop-ups colored the same way, until a new message is produced. If there is no new message during an entire typing session (a likely case), the color does not change at all. Then it is as if the keyboard were a regular one, not informative. In that sense, the informative keyboard output channel resembles the well-known Ambient Orb \cite{ambientWWWorb}, working at the ``change blind'' notification level \cite{matthews2004toolkit}. 

However, unlike in Ambient Orb, the color representing the current message is not displayed continuously; instead, the user receives multiple ``pulses'' (pop-ups). The pulses involve rapid changes (a pop-up appears and disappears), not generated autonomously by the display but brought about by the user via keypresses. Therefore, the pulses should not be confused with so-called transitions \cite{matthews2004toolkit}, a typical way for a peripheral display to demand attention (as found in the notification levels ``make aware,'' ``interrupt,'' and ``demand action'' \cite{matthews2004toolkit}). Thus, we propose a new notification level, ``user-driven pulsed repeated delivery,'' a variety of change blind. 

To appreciate the uniqueness of the user-driven pulsed repeated delivery, consider an alternative to the informative keyboard. Assume that an ``informative icon'' is a part of the home screen of the smartphone, and that its color represents a current message. The icon is then also a peripheral display, working at the generic ``change blind'' notification level. We hypothesize that user-driven pulsed repeated delivery makes it easier for messages to get noticed. To check the color of the icon, the user needs to focus on it, even if for a moment. The user of the keyboard receives multiple discrete ``reminders'' (pop-ups) of a single message, regardless of whether they intend to receive them or not. Verifying the hypothesis requires insight from cognitive science and is beyond the scope of this paper.

The notification level of a peripheral display is strongly related to its intrusiveness \cite{shelton2021gauging}. The informative keyboard does not proactively notify or interrupt the user. As argued above, it uses a combination of the (modified) ``ignore'' and ``change blind'' notification levels, which are the least distractive ones. The low notification levels suggest a low level of intrusiveness.

What makes the issue of intrusiveness less clear is that message delivery is bundled with typing -- the unrelated, important primary task that should be performed efficiently. During a typing session, the color of pop-ups changes rarely (if at all). However, the question remains whether typing performance is not impaired just by the fact that the color represents information. The intrusiveness of the informative keyboard and especially its impact on typing efficiency need to be determined experimentally.


\subsection{Informative keyboard as persuasive technology}

We have presented the informative keyboard output channel as a ``general purpose'' output device. However, our focus is on applying it to raise awareness of the amount of smartphone use. If messages contain feedback about problematic behavior, then the informative keyboard can be considered a persuasive technology, i.e., an interactive computing system designed to change user attitudes and behaviors \cite{fogg2003persuasion}, \cite{oinas2009persuasive}.

Providing feedback on behavior is a major technique used by persuasive technologies \cite{hermsen2016using}. Feedback can help users stop an undesirable behavior ``here and now'' or affect an underlying habit. To affect a behavior here and now, the feedback should be concurrent with the behavior. This allows ``reflection-in-action'' (when the behavior occurs), as opposed to ``reflection-on-action'' (when the behavior belongs to the past) \cite{hermsen2016using}. The informative keyboard, with its quasi-real time message delivery, supports reflection-in-action. The keyboard may remain unused for some hours at a time, but a typical user still receives the feedback several times a day and has quite a few opportunities to put the device aside. Thus, monitoring and control occur within the time frame of a day, a natural unit for these purposes. In addition, reflection-in-action is facilitated, as the feedback on the problematic behavior (i.e., using a smartphone) is embedded in the behavior itself. 

Some persuasive technologies start with the same premise as the informative keyboard, namely that detailed feedback may lead to a high cognitive load. As stated in \cite{ham2010ambient}, ``\ldots\ people often lack motivation or cognitive capacity to consciously process \ldots\ relatively complex information.'' In \cite{ham2010ambient}, this realization gives rise to the delivery of feedback via ambient (off screen) lighting. While the user sets a thermostat, the resulting amount of energy to be consumed is communicated through the color of the light. This is compared with ``factual feedback,'' displayed on a screen as numbers. It turns out that the lighting feedback leads to lower energy consumption (a stronger persuasive effect) and is easier to process \cite{ham2010ambient}. Although the informative keyboard is screen-based and differs otherwise, both solutions deliver minimal, color-encoded feedback. This is in line with a general design strategy put forward in the area of persuasive technologies, namely to use abstract data representation rather than ``raw and explicit data'' \cite{consolvo2009theory}.


%% file: heating_up.tex
\section{Heating up keyboard: informative keyboard to raise awareness of smartphone use} \label{section:heating}

\begin{figure}[t]
\centering
\includegraphics[scale=0.7]{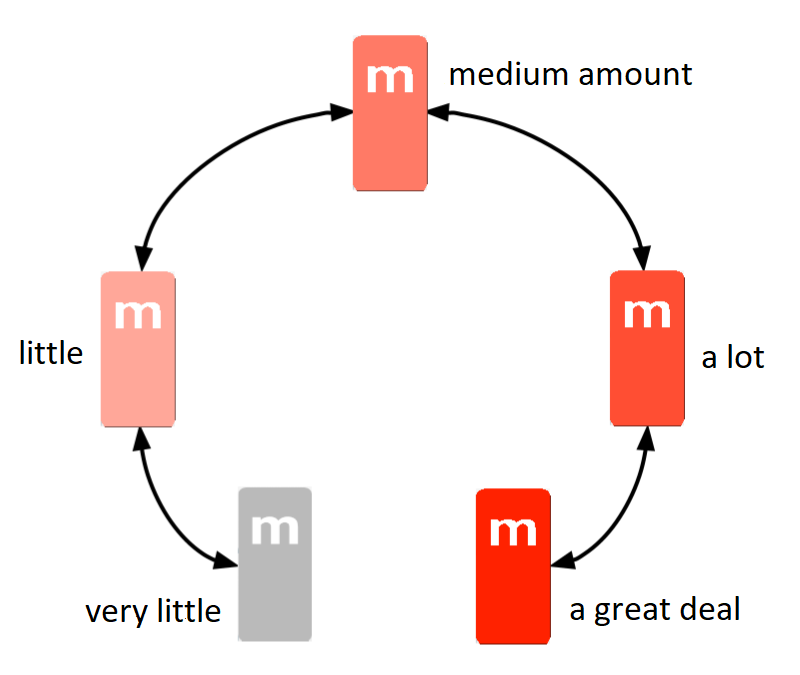}
\caption{The heating up keyboard provides feedback on the amount of smartphone use via red pop-ups of different saturation, as in heatmaps. The underlying feedback is referred to as the keyboard’s “temperature.”}
\label{fig:heating}
\end{figure}

We now focus on the informative keyboard that provides feedback on the amount of smartphone use. We start by assuming a specific palette of pop-up colors, which, while not inherent in the design, gives rise to a convenient analogy from physics. The palette, shown in Figure \ref{fig:heating}, consists of red colors with increasing saturation. The usual symbolism associated with red fits the nature of the information to be provided. The red color often conveys some negative state, danger, or a high value of a quantity, most notably temperature. As such, the color is commonly used in heatmaps. Accordingly, the red-based palette inspires us to refer to a message delivered by the keyboard as its ``temperature.'' We can thus say that the keyboard ``heats up'' or ``cools down,'' depending on whether a smartphone is used or not. The temperature-based phrasing seems intuitive when explaining the keyboard’s operation to the user. From now on, when discussing feedback on the amount of smartphone use, we will use the terms ``informative keyboard'' and ``heating up keyboard'' interchangeably.

Having assumed a specific palette, we note that it may have disadvantages (e.g., the red colors may be too intrusive) and can be replaced with another one, preferably by the user themselves.

In the following subsections, we present a baseline proof-of-concept feedback producing algorithm (i.e., an information source that feeds the heating up keyboard). Notably, just as the red-based palette is not the only choice, the presented algorithm is not either; there are many ways to conceptualize and estimate the amount of smartphone use. We also briefly describe our implementation of the entire system. 


\subsection{Estimating amount of smartphone use: selected design choices}

To estimate the amount of smartphone use, one needs to pick smartphone-based activities that should count. This is not easy, considering the vast diversity of possible activities and reasons users engage in them. Then, one needs to decide how the user’s smartphone-related behavior should be expressed through metrics, how often these metrics should be captured, how they should be combined into a consolidated estimate, and how the estimate should be mapped to the temperature of the heating up keyboard. 

In particular, one should decide on the ``memory'' of the algorithm. In our formulation, memory is described in terms of a time window and the speed of forgetting. The current temperature depends on the behavior of the user in a time window, which ends roughly``now.'' The window can be fixed in size and sliding, or it could be expanding, e.g., starting every day at midnight. As explained in the following, the window could even keep expanding without a limit. For a fixed-size window, one needs to choose its length. For a long window or one that can get long, the algorithm should probably include gradual forgetting of the past: what the user did several hours ago should matter less than what they did just a moment ago. With our model of memory, if the time window is long, and the forgetting is slow (a case of ``good memory''), it probably takes quite some time for the temperature to change in response to a change in behavior. 

As to the estimate-to-temperature mapping, one needs to choose the number of available temperature levels. One should also decide on the strictness (or leniency) of the algorithm, i.e., to pick thresholds that define the mapping, and thus values of the consolidated estimate that result in given temperature levels.


\subsection{Estimating amount of smartphone use: algorithm}

Regarding activities that count as smartphone use, we exclude using the device as an old-fashioned phone; talking on the phone does not count. We include two broad categories of activities: (a) browsing the Internet or using apps, and (b) checking notifications. Notifications are set apart because of their ability to disrupt whatever the user is doing. We do not differentiate between activities in either of the two categories. Simple metrics for them would be the time spent browsing or using apps and the number of notification checks, respectively.

For simplicity, we use only one metric, captured in discrete time, with a sampling period $T_S$. At the end of a sampling period, we consider so-called active intervals that occurred during that period. These are the times when the screen is active. We do not include the time when active intervals overlap with phone calls. If the duration of an active interval is quite short (below a threshold), we consider the interval to be due to a notification. In such a case, we calculate the interval’s corrected duration as $t_i^\prime = \max(t_i , T_N)$, where $t_i$ is its actual duration, and $T_N$ is the notification correction time. The correction reflects the disruptive impact of notification checks and other very brief interactions: they always count as at least the time $T_N$. If an active interval is longer than the threshold, then $t_i^\prime = t_i$.

After the corrections, the total usage time for a given sampling period is $t_u = \sum_i t_i^\prime$ (the sum includes all active intervals). We then define the period usage factor, $u = \min(t_u/T_S,1)$, as our metric. The correction may result in $t_u > T_S$, so we ensure that the period usage factor ranges from zero (no usage during the sampling period) to one (maximum usage). 

As can be seen, favoring simplicity over accuracy, we equate an active screen with actual smartphone use. This is only partially true. In many cases, the screen is active, but the user is not interacting with the smartphone. For example, a notification causes the screen to be active for some time, even if the user has put the smartphone away. However, one can argue that a high rate of notifications, even if some of them are not immediately attended to, is an indication of heavy smartphone use.

Now, let  $u_n$ be the period usage factor for the $n$th sampling period. We let the discrete-time signal $<u_n>$ be the input of an auto-regressive (AR) filter, $y_n=(1-\alpha)y_{n-1}+\alpha u_n$, where $0<\alpha<1$. We refer to the output sample $y_n$ as the overall usage factor at the time instant $n$ (at the end of the $n$th sampling period). Clearly, $0 \leq y_n \leq 1$, for all $n$. The overall usage factor is our consolidated estimate of the amount of smartphone use.

The overall usage factor summarizes usage over multiple sampling periods; in fact, sampling periods arbitrarily distant in time do contribute (i.e., the time window of our algorithm grows arbitrarily long). However, they contribute with decreasing weights; at the time instant $n$, the weights for $u_n$, $u_{n-1}$, $\ldots$, $u_{n-k}$, $\ldots$, are $\alpha$, $(1-\alpha)\alpha$, \ldots, $(1-\alpha)^k \alpha$, $\ldots$, respectively. For $\alpha$ close to one, previous periods quickly become irrelevant (fast forgetting). For $\alpha$ close to zero, the impact of past sampling periods decreases slowly. We refer to $\alpha$ and the filter as the forgetting coefficient and the forgetting filter, respectively.

The overall usage factor, $y_n$, can take on any value in the interval $[0,1]$. A quantizer is needed to map $y_n$ to one of few available temperature levels (messages). For example, if a source alphabet includes five messages, the quantizer partitions the interval $[0,1]$ into five quantization intervals. A uniform quantizer has quantization intervals of equal length: $[0,0.2)$, $[0.2,0.4)$, $\ldots$, $[0.8,1.0]$. In a non-uniform quantizer, quantization intervals have different lengths, which allows one to adjust the algorithm’s strictness (or leniency). For example, consider a quantizer with the quantization intervals such that the next one is twice as long as the previous one: $[0,1/31)$, $[1/31,3/31)$, $[3/31,7/31)$, $[7/31,15/31)$, and $[15/31,1.0)$. It yields a strict algorithm, as it suffices for $y_n$ to reach the value of about one-half for the highest temperature level to be displayed.

To approach strictness systematically, we introduce a strictness coefficient $s$ and use the function $f_s (x)=x^s$ to obtain the endpoints of the quantization intervals; they are $0.0$, $f_s (0.2)$, $f_s (0.4)$, $f_s (0.6)$, $f_s (0.8)$, and $1.0$. Consider the three values of $s$: $0.5$, $1.0$, and $2.0$. For $s=0.5$, the algorithm is lenient, with the successive quantization intervals getting shorter; the last one is (roughly) $[0.9,1.0]$. For s=2, the algorithm is strict: the quantization intervals become longer, and the last is $[0.64,1.0]$. For $s=1$, one gets the uniform quantizer and a ``neutral'' algorithm.

In summary, at the end of each sampling period, the algorithm adds up the durations of intervals when the screen is on (excluding those that coincide with phone calls), correcting the durations of quick notification checks. The time thus obtained is normalized and fed into a filter, which takes previous sampling periods into account, as prescribed by a forgetting coefficient. The output of the filter is quantized by a non-uniform quantizer, characterized by a strictness coefficient, to a temperature level displayed by the heating up keyboard during the next sampling period. The parameters, which can be adjusted (possibly by the user) to fine-tune our algorithm, are collected in Table 1.

The algorithm apparently complies with some design strategies identified in \cite{consolvo2009theory}. For example, the feedback is comprehensive in that most smartphone-based activities contribute to the keyboard's temperature. Also, the feedback represents a trend in the user's behavior: the memory-related mechanisms make the temperature a summary of the behavior in the (short-term) past. Finally, the feedback may be positive; the temperature may go up, but also down, several times a day. Nevertheless, both the algorithm and its parameters are based mainly on intuition and ease of implementation. A more elaborate design should be informed by psychological principles, research on smartphone overuse and information overload, as well as by experiments with users.


\begin{table}
\caption{Parameters of our smartphone usage estimation algorithm.}
\centering

{\footnotesize
\begin{tabular}{ p{0.23\textwidth} p{0.50\textwidth} p{0.15\textwidth} }
\midrule \midrule
\multicolumn{1}{c}{\textbf{parameter}} &
\multicolumn{1}{c}{\textbf{interpretation}} &
\multicolumn{1}{c}{\textbf{reasonable values}} \\
\midrule\midrule
sampling period $T_S$ & how often the amount of smartphone usage is sampled & 
\multicolumn{1}{r}{30 minutes} \\
\midrule
notification correction time $T_N$ & usage “time” recorded for very short interactions (e.g., notification checking), introduced to account for disruption & \multicolumn{1}{r}{5 minutes}\\
\midrule
forgetting coefficient \newline $\alpha$ & how quickly a past sampling period becomes less impactful when calculating the overall usage factor (the consolidated usage estimate) & \multicolumn{1}{r}{$0<\alpha<1$}\\
\midrule
strictness coefficient \newline $s$ & how much the overall usage factor needs to rise for the keyboard’s temperature to reach a higher level & \multicolumn{1}{r}{0.5, 1.0, 2.0}\\
\midrule \midrule
\end{tabular}
}
\label{table:algorithm_parameters}
\end{table}


\subsection{Implementation of heating up keyboard}

We implemented a proof-of-concept demonstrator of the heating up keyboard. The implementation includes a smartphone usage estimation algorithm, which is interfaced, as an information source, to a modified real-life smartphone keyboard application
\cite{obirek2023mobile}, \cite{sienicki2023use}, \cite{banasik2024innovative}.

As a major starting point, we used AnySoftKeyboard, an open source keyboard for the Android operating system \cite{anysoftkeyboardWWWonly}. We were able to make ``surgical'' changes in the source code of AnySoftKeyboard to control the pop-up color for each keypress. In the file \texttt{KeyPreviewPopupWindow.java}, at the beginning of the method \texttt{showPopup()}, we added one line, in which we invoke \texttt{ mPopupWindow.getBackground().setColorFilter()}. There, we pass the desired color of the pop-up as an argument. The desired color is communicated to AnySoftKeyboard by a usage monitoring application. The usage monitor and the informative keybopard communicate via binary SMS messages. We find SMS-based communication convenient, as it allows us to easily place the usage monitor and the keyboard on different devices; this might be useful in some future use cases. Thus, the other change to AnySoftKeyboard was to add a broadcast receiver, which receives the SMS messages, retrieves a color from their payload, and makes the color publicly available. Except for pop-up colors, no features of AnySoftKeyboard are affected by our modifications.

As to the usage monitor, we stick to the algorithm described above, and use assorted facilities afforded by the Android operating system. For example, we register broadcast receivers to capture the \texttt{SCREEN\_ON} and \texttt{SCREEN\_OFF} events. Notably, the implementation requires a good understanding of the Android programming model and API, as the application should operate correctly despite different actions taken by both the operating system and the user. 

The current realization of the heating up keyboard demonstrates the feasibility of the concept. Currently, we are working to improve both the algorithmic part (especially to refine the estimation of smartphone usage) and the implementation part (mainly to improve maintainability and overall quality from the software engineering point of view). 


%% file: survey.tex
\section{Evaluation survey} \label{section:survey}

 We carried out a pilot investigation of how potential users react to the heating up keyboard. We recruited students enrolled in the course ``IoT and Network Programming'' at the Faculty of Electronics and Information Technology, Warsaw University of Technology, taught by the first author. The evaluation session was followed by an unrelated lecture that was not on the course syllabus but was within its scope. Thus, we could offer attendees some bonus points towards the grade, as an incentive. In an invitation email, we informed that participants would learn about ``an unusual user interface for a mobile device.'' The session was conducted online, through MS Teams. At the beginning, the participants were told that they would provide input via an anonymous online questionnaire and that sensitive personal data would not be collected. We would know who participated, but not who completed a specific questionnaire. Participation beyond that point implied informed consent.

 Sixty-seven students, out of 90 enrolled, took part. There were 12 women and 55 men (the number disparity between females and males is typical of our faculty). All of them were in their early twenties (almost all between 22 and 24 years old). As mentioned, they studied at an information technology faculty at a technical university. 

 The evaluation session started with a brief slide presentation. There, we mentioned issues related to smartphone overuse, showed screens of a representative usage monitoring application, with a considerable number of settings and a lot of output data, and explained the basic principle of the operation of the heating up keyboard. Next, we played a short demonstration film. It presented a ``typical'' day of a user of the heating up keyboard, from the morning to the night. In the morning, the keyboard, which had not been used at night, was not warm at all and did not differ from a regular keyboard. Some ``bursts'' of smartphone-related activities were assumed to occur during the day, and the keyboard temperature was shown to rise as a result. It was shown that when the user put the smartphone aside, the keyboard cooled down. The activities and the temperatures were seen along with the time of day, for viewers to grasp the underlying correlation.
 

\subsection{Questionnaire}

Then we asked the participants to complete the online evaluation questionnaire. The questionnaire starts with a brief reminder of how the informative keyboard works (Figure \ref{fig:how_it_works_basic}) and information on how to interpret the keyboard temperature (Figure \ref{fig:how_it_works_parameters}).

\begin{figure}[hb]
\begin{center}
\fbox{
\begin{minipage}{0.9\textwidth}
{\itshape {\small The heating up keyboard works just like a regular one, with one exception: the color of the pop-up on keypress depends on how much you have been using your smartphone recently. The more you use the smartphone, the more intensely red the keyboard becomes. 

The color saturation, called the keyboard’s ``temperature,'' can take on five levels. When you start using the smartphone after a long break, the keyboard is not warm at all: the color is grey, just like in a regular keyboard. If you keep using the smartphone intensely, the temperature gradually rises, until the color gets bright red. If the intensity of use decreases, the temperature goes down. 

This way, seeing the pop-up color while typing, you receive an up to date, though approximate, feedback about your smartphone use. The five temperature levels correspond to phrases: ``very little,'' ``little,'' ``medium amount,'' ``a lot,'' ``a great deal.''

}}
\end{minipage}
}
\caption{The evaluation questionnaire: basic information for participants on how the heating up keyboard works (slightly edited).}
\label{fig:how_it_works_basic}
\end{center}
\end{figure}

The background information is followed by questions. They are divided into five sections, each focusing on one issue: (1) the reception of non-numerical and approximate feedback, (2) the impact of feedback delivery on typing, (3) the impact of receiving feedback on the user behavior, (4) an overall assessment of the heating up keyboard, and (5) assorted improvement suggestions. The questions, along with the percentages of different answers to close-ended questions, are presented in Tables \ref{table:questionnaire_section1}-\ref{table:questionnaire_section4}. 

\begin{figure}[t!]
\begin{center}
\fbox{
\begin{minipage}{0.9\textwidth}
{\itshape {\small The information delivered by the keyboard is approximate. You learn about how intensely you use your smartphone, but without details, so that not too much of your attention is asked for.

What does a specific temperature level mean? Every half an hour, the keyboard estimates the intensity of your smartphone use over, say, a dozen or so hours. Doing so, the keyboard applies a forgetting policy: the more recent your smartphone activity, the more it weighs in calculating the estimate. 

You will be able to define your own meaning of the temperature levels by setting some parameters. A forgetting rate specifies how quickly the keyboard forgets about the past, and thresholds are used to map the estimated intensity to one of the five temperature levels.

Consider two examples. For some specific thresholds, if you spend with your smartphone the same percentage of the time {\upshape every} half an hour, the temperature levels will mean the following:
\begin{tabbing}
``very little'' \quad\quad\quad\= less than 20\%\\
``little'' \> from 20\% to 40\%\\
``medium amount'' \> from 40\% to 60\%\\
``a lot'' \> from 60\% to 80\%\\
``a great deal'' \> 80\% or more
\end{tabbing}
For a specific forgetting rate, the speed of heating up due to {\upshape uninterrupted} use of the smartphone after a long break will be as follows:
\begin{tabbing}
``very little'' \quad\quad\quad\=	right away\\
``little'' \> after half an hour\\
``medium amount'' \> after an hour and a half\\
``a lot'' \> after two hours and a half\\
``a great deal'' \> after four hours
\end{tabbing}
}}
\end{minipage}
} 
\caption{The evaluation questionnaire: information for participants on how to interpret the keyboard’s temperature (slightly edited).}
\label{fig:how_it_works_parameters}
\end{center}
\end{figure}


\subsection{Results}

We first present the answers to the close-ended questions, section by section, and then the answers to all the open-ended questions, grouped together (the latter answers are often hard to categorize). Section 1 (see Table \ref{table:questionnaire_section1}) covers the basic principle of the operation of the heating up keyboard. We ask if it is easy to understand (Q1), and how the participants feel about feedback being non-numerical and approximate (Q2, Q3). As many as 100\%, 79\%, and 67\% of the answers were favorable to the keyboard. Notably, for Q2 and Q3, cautious responses (``rather'') substantially outnumbered definite ones (``definitely''), in either the ``yes'' or ``no'' version. Overall, most of the participants embraced the simplicity and approximate nature of the feedback, even if cautiously.

\begin{table}[t]
\caption{The evaluation questionnaire: section 1. Answers to close-ended questions:
1 – ``definitely no,'' 2 – ``rather no than yes,'' 3 – ``hard to say,'' 4 – ``rather yes than no,'' 5 = ``definitely yes.'' The numbers are percentages; they may not add up to 100\% due to rounding.
}
\centering
{\footnotesize
\begin{tabularx}{\textwidth}{X p{0.58\textwidth}
>{\raggedleft\arraybackslash}X
|>{\raggedleft\arraybackslash}X
|>{\raggedleft\arraybackslash}X
|>{\raggedleft\arraybackslash}X
|>{\raggedleft\arraybackslash}X}
\midrule\midrule
\multicolumn{7}{p{0.90\textwidth} }{\textbf{SECTION 1:} what do you think about \textbf{the way feedback is presented} by the heating up keyboard?} \\
\midrule
\multicolumn{2}{c}{\textbf{questions}} &
\multicolumn{5}{c}{\textbf{answers}} \\
\cmidrule{3-7}
\multicolumn{2}{c}{} &
\multicolumn{1}{c}{1} &
\multicolumn{1}{c}{2} &
\multicolumn{1}{c}{3} &
\multicolumn{1}{c}{4} &
\multicolumn{1}{c}{5} \\
\midrule\midrule
Q1 & 
Have you had any problems with understanding the general principle of how the feedback is visualized by the keyboard? &
70 & 30 & - & - & - \\
\midrule
Q2 &
Do you like the fact that the keyboard delivers feedback without numbers? &
-& 12 & 9 & 51 & 28 \\
\midrule
Q3 &
Do you like the fact that feedback from the keyboard is approximate (limited to five levels)? &
3 &	16 & 13 & 42 & 25 \\
\midrule
Q4 &
\multicolumn{1}{p{0.58\textwidth}}{Please share your comments about non-numerical, approximate feedback delivery.} &
\multicolumn{5}{c}{\textit{open-ended}} 
\\
\midrule\midrule
\end{tabularx}
}
\label{table:questionnaire_section1}
\end{table}

Section 2 (see Table \ref{table:questionnaire_section2}) focuses on the coupling of feedback delivery with typing, and a possible impact of the former on the latter. For most questions, a substantial majority of the responses (60\% to 70\%) were favorable to the heating up keyboard, but, as before, cautious answers (“rather”) prevailed. As many as 73\% of the participants answered ``definitely no'' or ``rather no than yes'' when asked whether feedback delivery would slow down their typing (Q7). On the other hand, only 37\% of the participants answered ``definitely no'' or ``rather no than yes'' when asked if high temperature would negatively affect their mood while or after typing; 43\% gave the opposite answers (Q10). Generally, while many participants did not expect that feedback delivery would impair their typing, a possible impact remains a key issue for further investigation, as it could likely cause the rejection of the heating up keyboard. 


\begin{table}[t]
\caption{The evaluation questionnaire: section 2. Answers to close-ended questions:
1 – ``definitely no,'' 2 – ``rather no than yes,'' 3 – ``hard to say,'' 4 – ``rather yes than no,'' 5 = ``definitely yes.'' The numbers are percentages; they may not add up to 100\% due to rounding.
}
\centering
{\footnotesize
\begin{tabularx}{\textwidth}{X p{0.58\textwidth}
>{\raggedleft\arraybackslash}X
|>{\raggedleft\arraybackslash}X
|>{\raggedleft\arraybackslash}X
|>{\raggedleft\arraybackslash}X
|>{\raggedleft\arraybackslash}X}
\midrule\midrule
\multicolumn{7}{p{0.90\textwidth} }{\textbf{SECTION 2:} what do you think about \textbf{receiving feedback while typing}?}\\
\midrule
\multicolumn{2}{c}{\textbf{questions}} &
\multicolumn{5}{c}{\textbf{answers}} \\
\cmidrule{3-7}
\multicolumn{2}{c}{} &
\multicolumn{1}{c}{1} &
\multicolumn{1}{c}{2} &
\multicolumn{1}{c}{3} &
\multicolumn{1}{c}{4} &
\multicolumn{1}{c}{5} \\
\midrule\midrule
Q5 &
Do you like the fact that feedback is delivered while you do something else (i.e., type)? &
4 & 16 & 9 & 54 & 16 \\
\midrule
Q6 &
Do you like the fact that you receive feedback on your smartphone use \textit{whenever} you type? &
9 &	19 & 10 & 31 & 30 \\
\midrule
Q7 &
Would the fact that the keyboard heats up (i.e., changes the pop-up color) interfere with your typing? &
16 & 48 & 13 & 19 &	3 \\
\midrule
Q8 &
Would the fact that the keyboard heats up slow down your typing? &
24 & 49 & 9 & 16 & 1 \\
\midrule
Q9 &
Would the fact that the keyboard heats up distract you while typing or after typing? &
12 & 40 & 7 & 37 & 3 \\
\midrule
Q10 &
Would a high temperature level negatively affect your mood while typing or after typing? &
7 & 30 & 19 & 28 & 15 \\
\midrule
Q11 &
Could you use the heating up keyboard even if you were not interested in feedback delivered by the keyboard (i.e., ignoring its temperature)? &
10 & 16 & 10 & 46 & 16 \\
 \midrule
 Q12 &
\multicolumn{1}{p{0.58\textwidth}}{Please share your comments about non-numerical, approximate feedback delivery.} &
\multicolumn{5}{c}{\textit{open-ended}} \\
\midrule\midrule
\end{tabularx}
}
\label{table:questionnaire_section2}
\end{table}

In Section 3 (see Table \ref{table:questionnaire_section3}) we inquire if the feedback provided by the keyboard would be relevant to the respondents, and if it would affect their behavior. As many as 73\% and 74\% answered ``definitely yes'' or ``rather yes than no'' when asked if awareness of the amount of smartphone use was important to them (Q13), and if the keyboard would help them become more aware (Q14). Counting the same way, fewer participants, but still a majority (56\%), declared that they would be inclined to resolve not to exceed a certain temperature level (Q17). The other questions in this section were answered similarly, and, again, cautious answers were more common.  Apparently, most participants implicitly acknowledge the problem of smartphone overuse (by wishing to become more aware) and believe that the feedback provided by the heating up keyboard has a chance to make a difference.


\begin{table}[!t]
\caption{The evaluation questionnaire: section 3. Answers to close-ended questions:
1 – ``definitely no,'' 2 – ``rather no than yes,'' 3 – ``hard to say,'' 4 – ``rather yes than no,'' 5 = ``definitely yes.'' The numbers are percentages; they may not add up to 100\% due to rounding.}
\centering
{\footnotesize
\begin{tabularx}{\textwidth}{X p{0.58\textwidth}
>{\raggedleft\arraybackslash}X
|>{\raggedleft\arraybackslash}X
|>{\raggedleft\arraybackslash}X
|>{\raggedleft\arraybackslash}X
|>{\raggedleft\arraybackslash}X}
\midrule\midrule
\multicolumn{7}{p{0.90\textwidth} }{\textbf{SECTION 3}: would feedback delivered by the heating up keyboard \textbf{be relevant to you/affect you}?}\\
\midrule
\multicolumn{2}{c}{\textbf{questions}} &
\multicolumn{5}{c}{\textbf{answers}} \\
\cmidrule{3-7}
\multicolumn{2}{c}{} &
\multicolumn{1}{c}{1} &
\multicolumn{1}{c}{2} &
\multicolumn{1}{c}{3} &
\multicolumn{1}{c}{4} &
\multicolumn{1}{c}{5} \\
\midrule\midrule
Q13 &
Is it important to you to be aware how much you use your smartphone? &
4 & 18 & 4 & 39 & 34 \\
\midrule
Q14 &
Could the keyboard help you become aware how much you use your smartphone? &
4 & 10 & 10 & 43 & 31 \\
\midrule
Q15 &
Would you feel uncomfortable seeing the keyboard’s temperature rising to a high level? &
4 & 19 & 6 & 43 & 27 \\
\midrule
Q16 &
Would you be inclined to put your smartphone aside if you saw that the keyboard’s temperature has risen to a high level? &
1 & 19 & 7 & 58 & 13 \\
\midrule
Q17 &
Would you be inclined to make a resolution not to exceed a certain temperature level? &
12 & 22 & 9 & 40 & 16 \\
\midrule
Q18 &
Would the fact that the keyboard regularly heats up to high temperatures motivate you to reduce using your smartphone? &
3 & 12 & 13 & 45 & 27 \\
\midrule
Q19 &
Could regular reception of the keyboard’s temperature affect your smartphone use habits? &
3 & 16 & 18 & 49 & 13 \\
 \midrule
Q20 &
\multicolumn{1}{p{0.58\textwidth}}{Please share your comments as to whether the heating up keyboard can help one to use one’s smartphone with more awareness.} &
\multicolumn{5}{c}{\textit{open-ended}} \\
\midrule\midrule
\end{tabularx}
}
\label{table:questionnaire_section3}
\end{table}

Section 4 (see Table \ref{table:questionnaire_section4}) is about an overall assessment of the heating up keyboard. Around 60\% answered ``definitely yes'' or ``rather yes than no'' when asked if they would recommend the keyboard to friends and family (Q21). However, only 43\% (still a majority) answered this way when asked if they would like their keyboard to have the option to heat up (Q23). 


\begin{table}[t]
\caption{The evaluation questionnaire: sections 4 \& 5. Answers to close-ended questions:
1 – ``definitely no,'' 2 – ``rather no than yes,'' 3 – ``hard to say,'' 4 – ``rather yes than no,'' 5 = ``definitely yes.'' The numbers are percentages; they may not add up to 100\% due to rounding.}
\label{table:questionnaire_section4}
\centering
{\footnotesize
\begin{tabularx}{\textwidth}{X p{0.58\textwidth}
>{\raggedleft\arraybackslash}X
|>{\raggedleft\arraybackslash}X
|>{\raggedleft\arraybackslash}X
|>{\raggedleft\arraybackslash}X
|>{\raggedleft\arraybackslash}X}
\midrule\midrule
\multicolumn{7}{p{0.90\textwidth} }{\textbf{SECTIONS 4}: please tell us about \textbf{your overall assessment} of the heating up keyboard (Q21-24). \textbf{SECTIONS 5}: please \textbf{help us design} a good heating up keyboard (Q25).}\\
\midrule
\multicolumn{2}{c}{\textbf{questions}} &
\multicolumn{5}{c}{\textbf{answers}} \\
\cmidrule{3-7}
\multicolumn{2}{c}{} &
\multicolumn{1}{c}{1} &
\multicolumn{1}{c}{2} &
\multicolumn{1}{c}{3} &
\multicolumn{1}{c}{4} &
\multicolumn{1}{c}{5} \\
\midrule\midrule
Q21 &
Would you recommend the heating up keyboard to your friends or family? &
1 & 12 & 25 & 51 & 10 \\
\midrule
Q22 &
Would you like the smartphone keyboard that you currently use to have the option to heat up? &
4 & 24 & 28 & 28 & 15 \\
\midrule
Q23 &
Is adding to a keyboard the capability to heat up a good solution? &
- & 4 & 19 & 60 & 16 \\
\midrule
Q24 &
\multicolumn{1}{p{0.58\textwidth}}{Please share your comments about your overall assessment of the heating up keyboard.} &
\multicolumn{5}{c}{\textit{open-ended}} \\
\midrule
Q25 &
\multicolumn{1}{p{0.58\textwidth}}{Please share your comments and ideas as to what to improve in the way the heating up keyboard works. What would you change in the presented keyboard?} &
\multicolumn{5}{c}{\textit{open-ended}} \\
\midrule\midrule
\end{tabularx}
}
\end{table}

The open-ended questions (Q4, Q12, Q20, Q24, Q25) yielded a plethora of comments, which we classify as follows: (a) comments praising the concept and/or its implementation, (b) constructive comments suggesting specific improvements, and (c) comments with reservations and criticism. Many participants offered several comments, often belonging to different categories.

There were many positive comments. Most of the participants appreciated visual (non-numerical) and approximate feedback. An example: ``Non-numerical representation is more readable.'' A more in-depth opinion, with a justification, is as follows: ``This is a very good approach, as we are often overloaded with graphs, numbers, percentages, etc. In this case, the user gets information represented visually, while using the device. It is a better solution, especially these days, when people have hard time focusing their attention.''  Another comment along the same lines reads: ``It is easier to notice a change when this is presented by color rather than numbers; at some point in time users would start to ignore numbers, no matter what they are.'' In the following comment, there is appreciation for implicit interaction as a means to provide feedback: ``This is a good solution, as we do not have to focus on analyzing numbers; we get the information at once, without making this our objective. As to other applications, one must start them to get the information; I think many people forget to do that after some time, or they simply do not feel like doing that.'' Another participant stated: ``I do not check usage statistics offered by my smartphone; but seeing changing colors, I would monitor my usage.'' A participant liked receiving feedback ``involuntarily''; another approved of the fact that information is displayed ``without asking.'' A couple of participants used the word ``subliminal'' to describe the impact of the keyboard on the user.

The second category of comments are those with a constructive suggestion on how to extend or modify the heating up keyboard without changing its model of user interaction. We went through all the open-ended questions and compiled a table with suggestions (see Table \ref{table:constructive}). For each suggestion, we counted how many times it occurred.


\begin{table}
\caption{The evaluation questionnaire: constructive suggestions extracted from answers to open-ended questions (Q4, Q12, Q20, Q24, Q25). The last column contains the number of occurrences of a comment; the number of participants was sixty-seven.
}
\centering

{\footnotesize
\begin{tabular}{ p{0.02\textwidth} p{0.74\textwidth} p{0.10\textwidth} }
\midrule\midrule
\multicolumn{1}{c}{\textbf{no.}} &
\multicolumn{1}{c}{\textbf{constructive suggestion}} &
\multicolumn{1}{c}{\textbf{occurred}} \\
\midrule\midrule
1. & 
Allow the user to choose a palette of colors for pop-ups on keypress.
Add comprehensive customization/personalization mechanisms. & 
\multicolumn{1}{r}{34/67}\\
\midrule
2. & 
Add a separate app allowing traditional inspection of numerical usage data. & 
\multicolumn{1}{r}{22/67}\\
\midrule
3. & 
Increase the number of temperature levels, make it configurable, or do not quantize. & 
\multicolumn{1}{r}{10/67}\\
\midrule
4. & 
Change the color of all the keys, or of some high-profile keys (e.g., the space key), 
or of the key being pressed itself (rather than the pop-up). & 
\multicolumn{1}{r}{7/67}\\
\midrule
5. & 
Complement color with notifications, sound, or vibration at selected temperature levels. & 
\multicolumn{1}{r}{7/67}\\
\midrule
6. &  
When estimating the amount of smartphone use, do not count purposeful, productive, or otherwise 
“justified” usage (e.g., using for work, using on an airplane, using to navigate, etc.). 
Such usage may be detected by the phone or marked by the user. & 
\multicolumn{1}{r}{5/67}\\
\midrule
7. &  
Clearly explain how the keyboard works. Explain what is meant numerically by different temperature levels. 
Also, explain that heating up does not occur only while typing. & 
\multicolumn{1}{r}{5/67}\\
\midrule
8. & 
Make it possible to easily switch temperature delivery on and off (e.g., add a button). & 
\multicolumn{1}{r}{4/67}\\
\midrule
9. & 
Do not force the user to change keyboards to get the functionality of heating up. & 
\multicolumn{1}{r}{4/67}\\
\midrule
10. &  
Add gamification (e.g., a reward for not exceeding some temperature levels). & 
\multicolumn{1}{r}{4/67}\\
\midrule
11. &  
Deliver feedback only every nth character, not with every character. & 
\multicolumn{1}{r}{1/67}\\
\midrule
12. & 
Do not use very bright/intense colors to avoid distraction/irritation. & 
\multicolumn{1}{r}{1/67}\\
\midrule\midrule
\end{tabular}
}
\label{table:constructive}
\end{table}

The most frequent suggestion was to allow the user to customize the palette of colors and, in general, to make the heating up keyboard widely configurable. The palette of green, yellow, and red was proposed several times. The second most common suggestion was to complement the keyboard with an add-on application that provides usage statistics in a common, numerical form. The third most popular suggestion was to increase the number of temperature levels, to make it configurable, or to skip the quantization altogether. Other, less frequent suggestions can be found in Table \ref{table:constructive}.

The third category of comments are those with reservations and criticism (see Table \ref{table:criticism}). By far the most frequent reservation was that users often spend a long time with the smartphone without using its keyboard, so they are left without feedback. Popular applications that offer infinite scroll (e.g., TikTok, Instagram) were frequently mentioned in this context. Quite a few comments expressed the concern that feedback delivery can disrupt, irritate, or negatively interfere with typing. (However, some of those concerned remarked that these negative kinds of impact may not be so bad, as they may result in putting the smartphone aside.) An (implicit) criticism was expressed by suggesting that feedback, although qualitative and approximate, should not be delivered with a keyboard but with another element of the user interface; an ``informative icon'' on the status bar was proposed several times. The less frequent critical comments can be found in Table \ref{table:criticism}. Notably, the comments in Table \ref{table:criticism} do not necessarily imply rejection of the heating up keyboard; often, they were contributed by participants who at the same time considered the concept worth pursuing.


\begin{table}[ht]
\caption{The evaluation questionnaire: reservations and criticism extracted from answers to open-ended questions.
(Q4, Q12, Q20, Q24, Q25). The last column contains the number of occurrences of a comment; the number of participants was sixty-seven.
}
\centering

{\footnotesize
\begin{tabular}{ p{0.02\textwidth} p{0.74\textwidth} p{0.10\textwidth} }
\midrule\midrule
\multicolumn{1}{c}{\textbf{no.}} &
\multicolumn{1}{c}{\textbf{comment with reservations or criticism}} &
\multicolumn{1}{c}{\textbf{occurred}} \\
\midrule\midrule
1. & 
The keyboard may remain unused for a long time even though the smartphone is being used 
(e.g., consider infinite scrolling). No feedback during those time intervals. & 
\multicolumn{1}{r}{22/67}\\
\midrule
2. & 
The heating up keyboard may be disrupting or irritating, or it may affect 
the speed of typing, especially if the smartphone is used for focused, productive, purposeful activities. & 
\multicolumn{1}{r}{18/67}\\
\midrule
3. & 
Deliver the temperature feedback with color (as in the heating up keyboard),
but apply the color to another UI element (e.g., an icon on the status bar or the borders of the screen). 
Alternatively, replace color with notifications at certain temperature levels. & 
\multicolumn{1}{r}{16/67}\\
\midrule
4. & 
The user may eventually get used to the temperature feedback and start to ignore it. & 
\multicolumn{1}{r}{11/67}\\
\midrule
5. & 
Some people simply prefer numbers (detailed numerical statistics), and/or ``old-fashioned'' numerical feedback-based applications would work just as effectively. &
\multicolumn{1}{r}{6/67}\\
\midrule
6. &  
Some people disable pop-ups on keypress. & 
\multicolumn{1}{r}{1/67}\\
\midrule
7. &  
Some people need strictly enforced limits (a brute force approach), not feedback. &
\multicolumn{1}{r}{1/67}\\
\midrule\midrule
\end{tabular}
}
\label{table:criticism}
\end{table}

Finally, we report a rather popular comment (with 10 occurrences), outside of the above categories: the heating up keyboard is a solution for those who do want to reduce their smartphone usage or become more aware. Those who are comfortable with the current usage will not find it attractive.


\subsection{Discussion}

Generally, the answers to the close-ended questions were favorable to the presented solution. At the same time, they were somewhat halfhearted: as a rule, cautious responses outnumbered definite ones. This might be caused by the fact that during the evaluation session the participants had not had hands-on experience with the heating up keyboard. At the very least, most of the participants did not exhibit a negative sentiment, which is encouraging. This is consistent with numerous positive comments offered in response to the open-ended questions.

The constructive suggestions (see Table \ref{table:constructive}) assume no change in the basic concept of the heating up keyboard; most are worthy of consideration. Several of them quite predictably call for customization mechanisms (Table \ref{table:constructive}, items 1, 3, 4, and 13), especially the ability to customize the palette of colors. The very popular suggestion to have a supplementary app (item 2) indicates that while non-numerical and approximate feedback was widely embraced, many participants would still like to have the option to access feedback as detailed, numerical data. The suggested feature of being able to switch the pop-up coloring off (item 8), and thus revert to a ``regular'' keyboard, would help at times when the feedback affects typing or productivity. The suggestion not to count certain uses of the smartphone (item 7) comes from the realization that some usage is ``legitimate.'' For example, using the smartphone for work should not cause the keyboard to heat up. The challenge is to seamlessly incorporate such exemptions without complicating the interaction model. Finally, there are suggestions, which propose that feedback be delivered somewhat differently, although still via the keyboard and coloring: to color all the keys of the keyboard rather than pop-ups (item 5), to complement pop-up coloring with, e.g., sound or vibration (item 6), or to color the pop-up for every $n$th entered character, with neutral pop-ups for remaining $n-1$ characters (item 12). Interestingly, the first two of these proposals are likely to make the heating up keyboard more intrusive, while the third one --- less intrusive.

We now turn to reservations and criticism (see Table \ref{table:criticism}), touching on the most frequently raised issues. Indeed, a considerable delay in receiving feedback can result when a smartphone is used without using its keyboard (item 1). However, as shown in \cite{druijff2021behavioural}, \cite{smolders2023smartphone}, typing is a frequent activity that occurs in most waking hours (so that the lack of typing indicates sleep). Even if occasional gaps in feedback delivery last for several hours, the user should still be able to control their smartphone usage on any given day. This is what we refer to as quasi-real time. 

Another concern is that the feedback provided by the heating up keyboard may be disturbing, or irritating, or may slow down typing (Table \ref{table:criticism}, item 2). This crucial issue deserves further investigation. However, the two just mentioned constructive suggestions (Table \ref{table:constructive}, items 8 and 12), to enable the user to switch pop-up coloring off and to color a pop-up only every $n$th character, may mitigate the problem.

The proposal (Table \ref{table:criticism}, item 3) that feedback, still non-numerical and approximate, be delivered by another element of the smartphone user interface (e.g., a colored icon on the status bar), should be investigated in separate research. In Section \ref{section:HCI}, we contrast an icon-based approach with the keyboard-based one. There, we claim that the informative keyboard introduces a new peripheral display notification level (user-driven pulsed repeated delivery), with possible merits.


%% file: related_work.tex
\section{Related work} \label{section:related_DSCT}

In this section, we mention some digital self-control tools (DSCTs) \cite{roffarello2023achieving}. We include DSCTs that target smartphones and in which we could find a quality reminiscent of the heating up keyboard: minimalism, non-textual feedback representation, delivery based on implicit interaction, or the lack of coercion. Often,  one or more of these qualities relate to only a tiny part of the overall functionality of a DSCT. 

In Good Vibrations \cite{okeke2018good}, the user is provided with two types of feedback. One is text-based and includes just two numbers: how long a target application has been used and the number of times it has been opened. This information is updated in real time, but to access it, the user has to swipe down on the status bar. The other feedback is delivered through vibration-based alerts that occur when the application is used beyond its usage limit. In both Good Vibrations and the heating up keyboard, the feedback is minimal. They both offer nonverbal feedback, which is accessed via implicit interaction. In Good Vibrations, the feedback is updated in real time, while the heating up keyboard delivers its feedback in quasi-real time. The authors describe their solution in terms of digital nudge and negative reinforcement \cite{okeke2018good}. The heating up keyboard can be described similarly: pop-ups indicating a high keyboard temperature may be considered an unwelcome nudge, which people would rather avoid. 

In Menthal \cite{andone2016menthal}, smartphone usage is quantified with one number, called MScore. It is a weighted average of four metrics: the time spent on a device, the number of times the device is unlocked, the number of SMS messages exchanged, and the time of phone calls. Its value, between 1 and 100, is available via a dedicated screen or a notification. Both Menthal and the heating up keyboard present the user with a single-value summary of their smartphone usage. The Menthal rationale for doing so (``A large part of the population find it difficult to \ldots\ to understand statistical data.'' \cite{andone2016menthal}) applies to our approach as well. However, we take the simplification a step further: we quantize an estimate of the amount of smartphone use into one of a few quantization levels. Also, we deliver this approximate value as the keyboard temperature using color, not text. Notably, there is a degree of interoperability between Menthal and the heating up keyboard. In Section \ref{section:heating} we presented a possible smartphone usage estimation algorithm but made a disclaimer that it can be replaced. If the algorithm producing MScore is extended by adding a quantizer, its output can be fed to the heating up keyboard. In fact, the heating up keyboard can be used as an output device with any algorithm that captures smartphone usage via a single number.

Habit Tracker \cite{shen2019development} captures the usage of applications in several categories and records offline user behaviors related to health. It presents the captured data in detail. However, it also combines the data and summarizes a day with a single number, called the daily achievement score. The score is then used as the driving force behind the growth of a ``tree.'' The higher the achievement score, the faster the tree grows. The user inspects a picture of the tree to see how they are doing. The heating up keyboard also works with a single number summary and also offers a non-verbal representation of the summary. Our representation is more abstract, as it does not involve any object.

An approach similar to that of Habit Tracker is applied in the popular productivity app called Forest \cite{seekrtechWWWforest}. Using Forest is a sequence of challenges: each consists in setting a timer and declaring that the smartphone will not be used for the specified time. When the timer starts, a virtual tree is ``planted'' and continues to grow as long as the smartphone is not being used. If the smartphone remains unused until the timer expires, the tree is healthy; otherwise, it dies. In either case, the plant, alive or dead, is added to a virtual forest. The forest is a pictorial representation of the user's longer-term behavior. Both Forest and the heating up keyboard convey information in a non-verbal way.

In ScreenLife \cite{rooksby2016personal}, total screen time is the metric captured and made available to users. The screen time is presented mostly nonverbally, through colored bars, with color saturation increasing with the amount of usage per hour. These are similarities, but the heating up keyboard also differs significantly: to access information on usage, one does not have to open an app but simply to type anything. This makes it more likely that the user will be exposed to feedback. Participants in an in-the-wild study presented in \cite{rooksby2016personal} often did not open ScreenLife for many days in a row (except during the first couple of days of the experiment). Another difference, related to our proposed usage estimation algorithm, is that the temperature level of the heating up keyboard can go up and down during the day, reflecting recent user behavior; the total screen time can only go up. 

%% file: summary.tex
\section{Summary and future work} \label{section:summary}

In this paper, we introduced the informative keyboard as a ``general-purpose'' smartphone output device. Next, we focused on using it to increase awareness of the amount of smartphone use (and we adopted the more specific term ``heating up keyboard''). We presented the concept and an implementation of the heating up keyboard to a group of potential users. Their opinions indicate that the approach may be embraced by quite a few people. We also received constructive suggestions, as well as comments with reservations and criticism. 

We see three directions for further exploration: (a) experimenting with the heating up keyboard in the wild, (b) improving its current design, and (c) exploring other uses of the informative keyboard.  

An experiment, in which the heating up keyboard is used for an extended period, should address its usability and utility. The usability has to do with any potential impact of the keyboard’s output channel on typing, the key functionality that should not be affected. Is delivering messages intrusive? Does it contribute to cognitive load or slows down typing? Does it cause discomfort (e.g., due to receiving a piece of ``unpleasant'' feedback). If so, are these sufficient reasons to abandon the heating up keyboard? On the other hand, do users like the fact that the interaction occurs even if it is not intended? The utility issue is whether the heating up keyboard can do the job it was designed to do: to raise awareness and, possibly, to lead to a reduction of the amount of smartphone use. 

There are many potential improvement areas: customization, explainability, intrusiveness, estimation of the amount of smartphone use, context awareness, and modalities other than color. Customization and explainability can be addressed by a companion app. The app should allow one to choose the palette of colors and configure the algorithm that estimates the amount of smartphone use. The app should also offer help with the meaning of colors and not-so-obvious algorithm parameters (like our forgetting coefficient). A keyboard temperature level, while concise and easy to observe, may be confusing to the user, who may ask how their behavior has resulted in the particular level \cite{andone2016menthal}. Therefore, the companion app should make raw usage data available to the user and offer an explanation on how a temperature level was derived from the data. The companion app could also enable the user to reset the usage estimation algorithm, which would instantly reduce the keyboard temperature to its minimum value; this would follow the design strategy to make persuasive technology controllable \cite{consolvo2009theory}.

Another key area is intrusiveness; one would like to have mechanisms to adjust the ``intensity'' of keyboard-generated feedback. The constructive suggestion that the current temperature level should be delivered every $n$th character can be followed in different ways. For example, it could be understood deterministically or in terms of an average. Also, the parameter $n$ can be set by the user, or it can depend on the current temperature level: as the temperature increases, $n$ decreases, so the temperature is shown more frequently. Then, more intense smartphone usage would be signaled by both the pop-up color for selected keypresses and by how often such keypresses occur.

An approach to managing the intrusiveness is to allow pop-up coloring to be completely switched off. Switching it back on could be done by the user or could happen automatically after some time, to prevent it from being permanently disabled. The logic behind heating up should work all the time, even if feedback delivery is switched off (so that the keyboard temperature always reflects actual usage).

To take the reduction of feedback intensity to an extreme, one can experiment with a heating up keyboard that delivers subliminal feedback: a color representing the current temperature level could be applied for a very short interval (in the order of milliseconds); after that, the pop-up would be colored neutrally \cite{ham2009can}.

A major issue is how to make a single-number estimate truly indicative of the impact of the smartphone on its owner. Screen time is not the only metric; the frequency of smartphone checks may be more relevant \cite{singaporeWWWfrequent}. Some uses are purposeful and productive, so they should not be counted. Different activities exert different levels of cognitive load on the user. Finally, the user may only be interested in some specific activities, e.g., using a selected application. These factors should be taken into account when designing a usage estimation algorithm and the companion add with its customization options. 

Implicit interaction often relies on context, which is acquired by the system without explicit user input. The context could be used to detect productive uses. For example, if a time and location indicate that the user is at work, smartphone usage may not be counted, and feedback may not be delivered. Or, the system can detect the purpose of typing to enable message delivery only for less important tasks (e.g., entering search phrases) and to disable it for more important typing (e.g., composing text messages). 

In this paper, the medium used to deliver messages is the background color of pop-ups on keypress. However, one can experiment with other modalities of keypress-coupled feedback. In an ``auditory informative keyboard,'' a message would be delivered via its unique key-click sound. In the ``haptic informative keyboard,'' each message would have its own vibration ``signature.'' A lot of what has been said on pop-ups applies to the other two modalities, but the overall user experience would have to be investigated separately for each of them.

The third direction of future work is to explore whether the heating up keyboard (and, more generally, the informative keyboard) can be used differently from the way described here. For example, imagine applying the heating up keyboard in a soft ``parental awareness system,'' where the temperature of the parent’s keyboard depends on how much a smartphone is used by the child. In Section 2, we present the keyboard’s output channel as a ``general purpose'' output device and mention some other sources of information that can be connected to it. More examples of such sources are likely to be found.